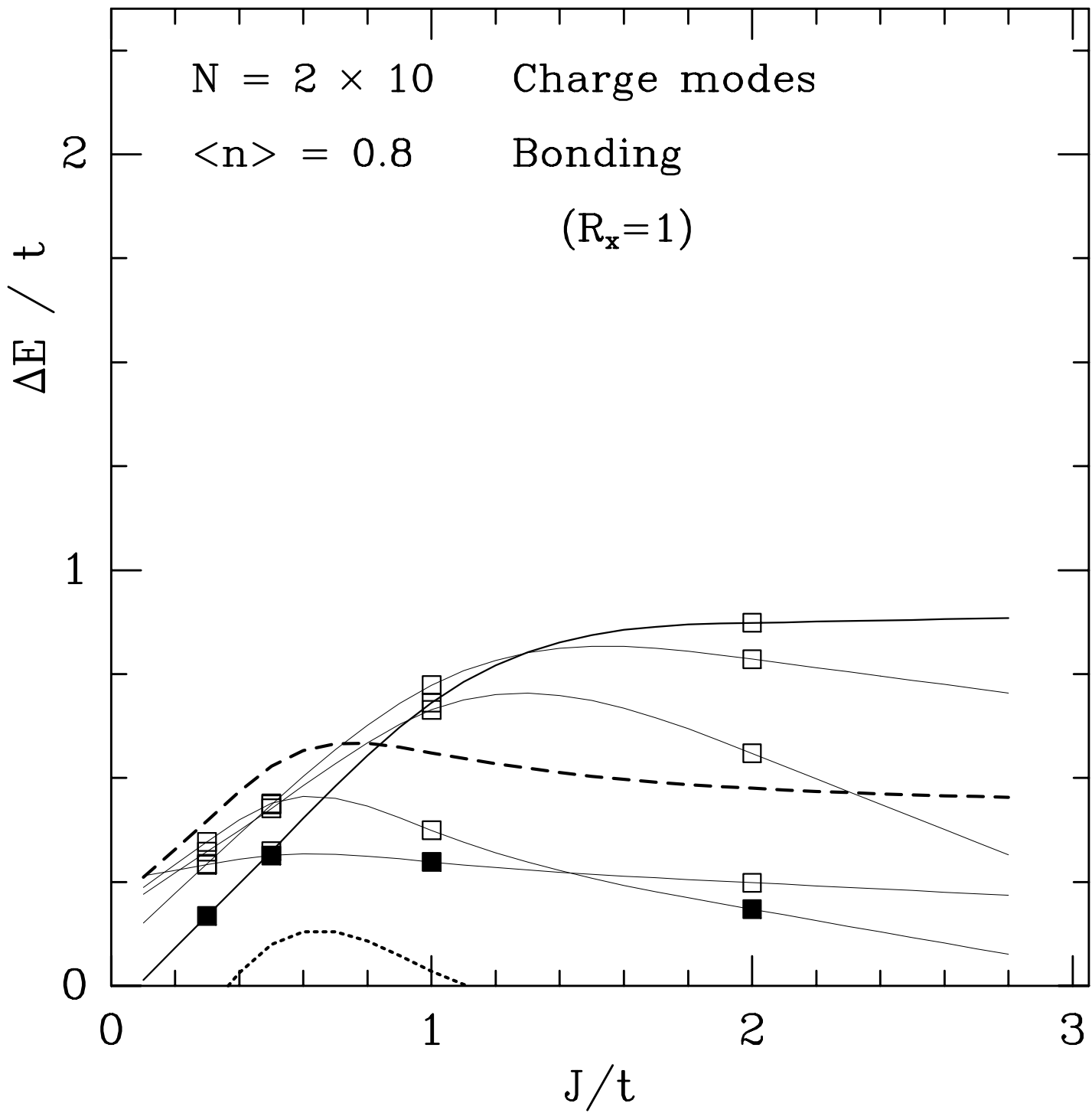

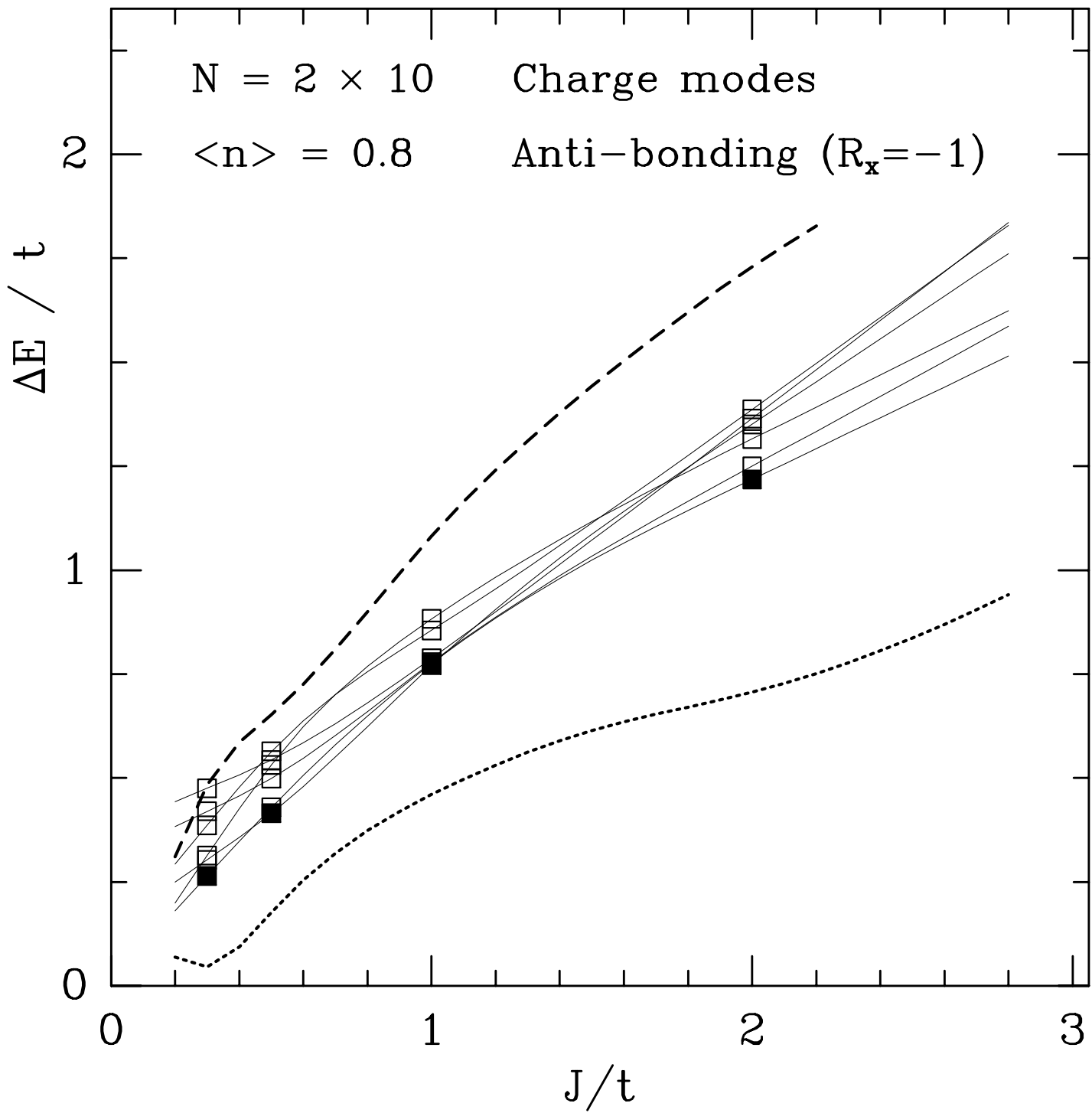

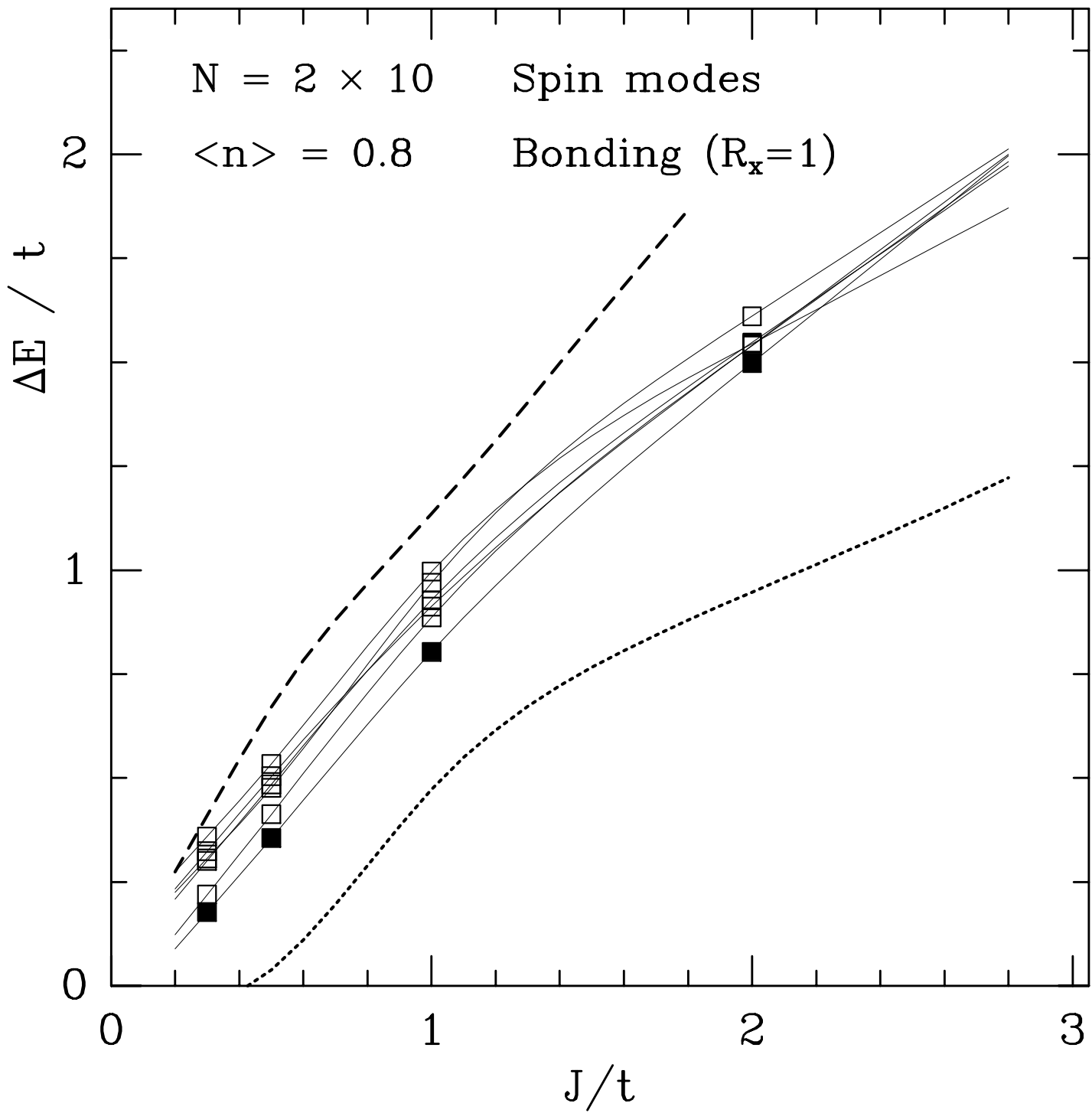

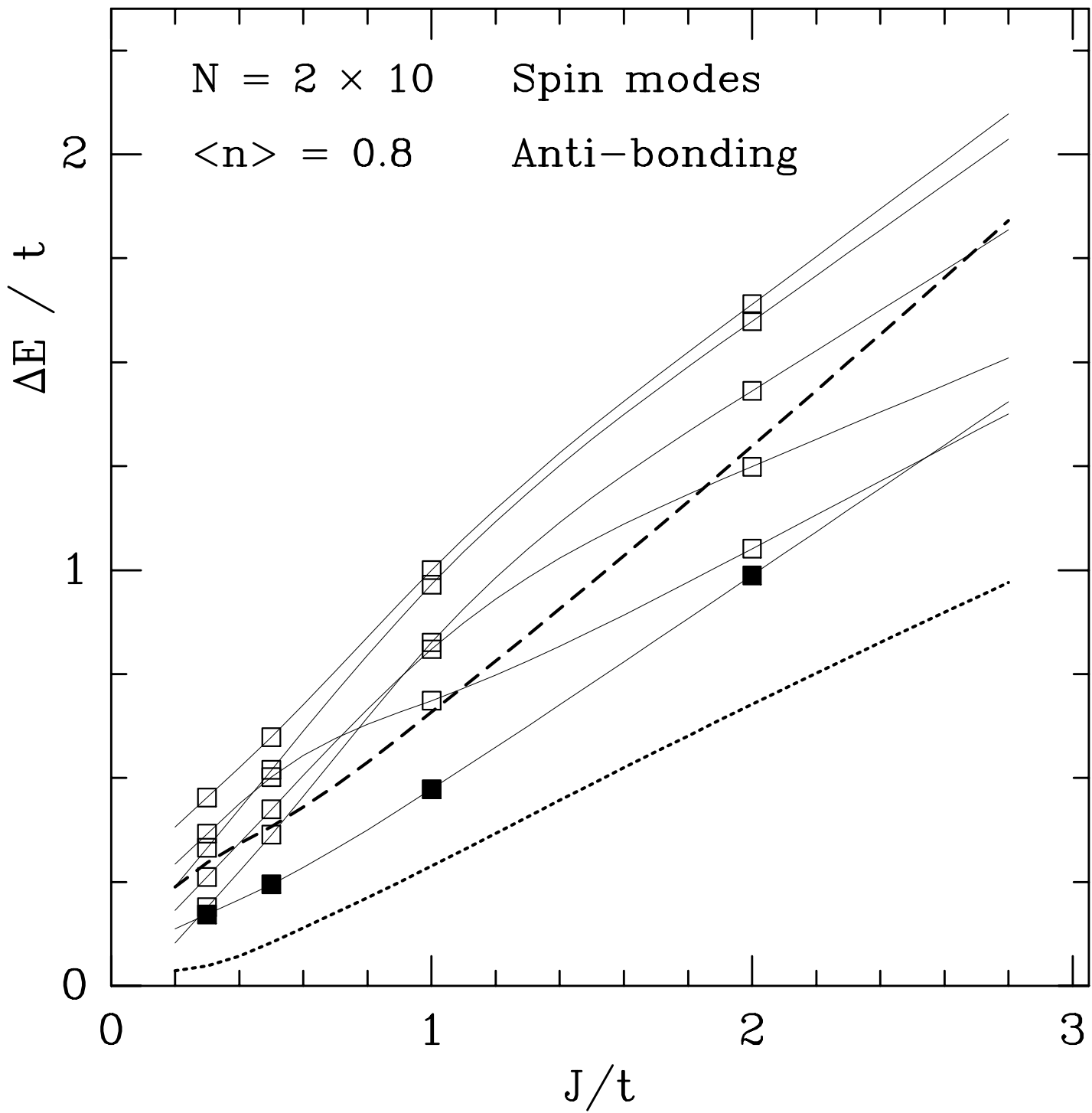

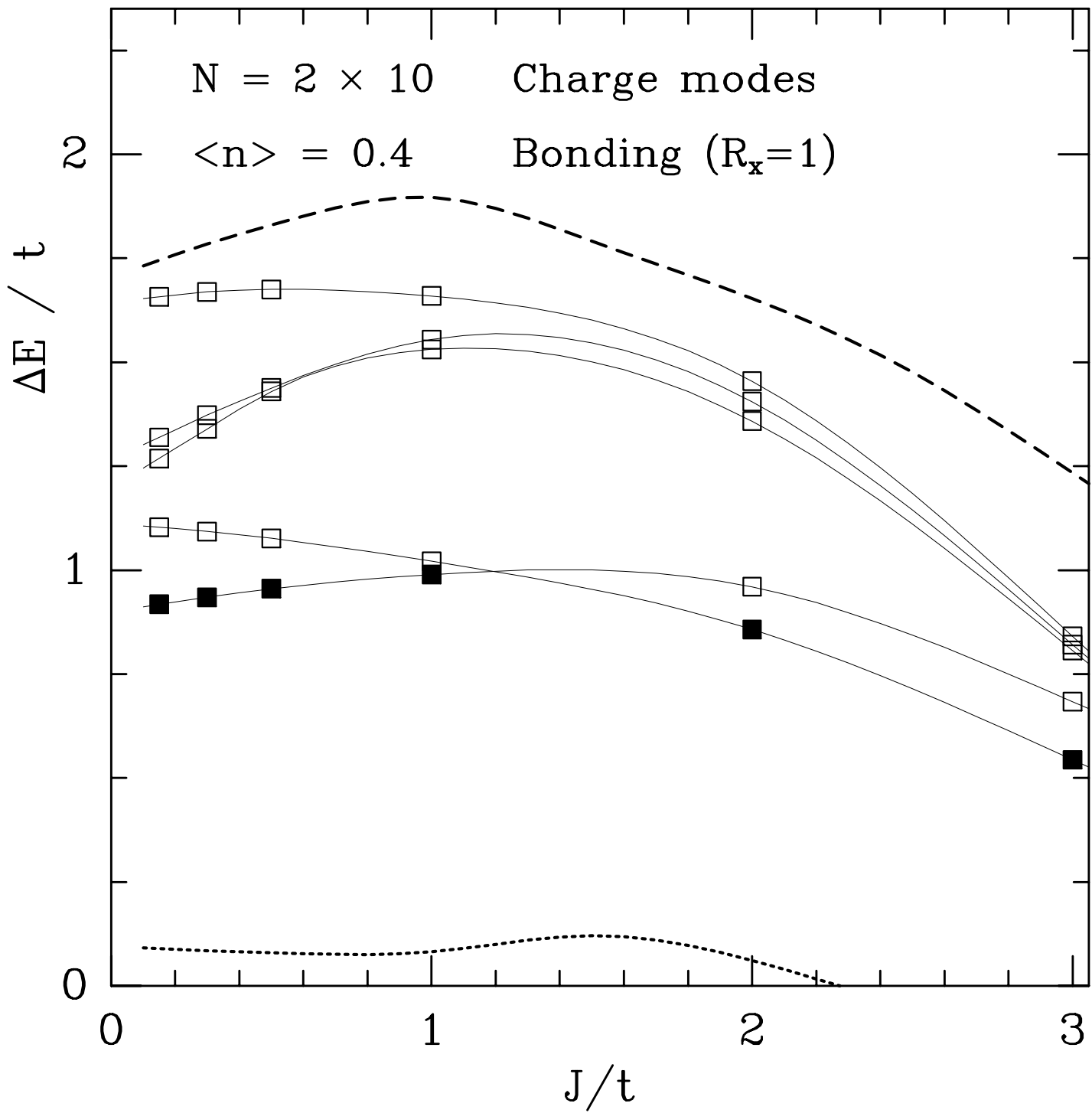

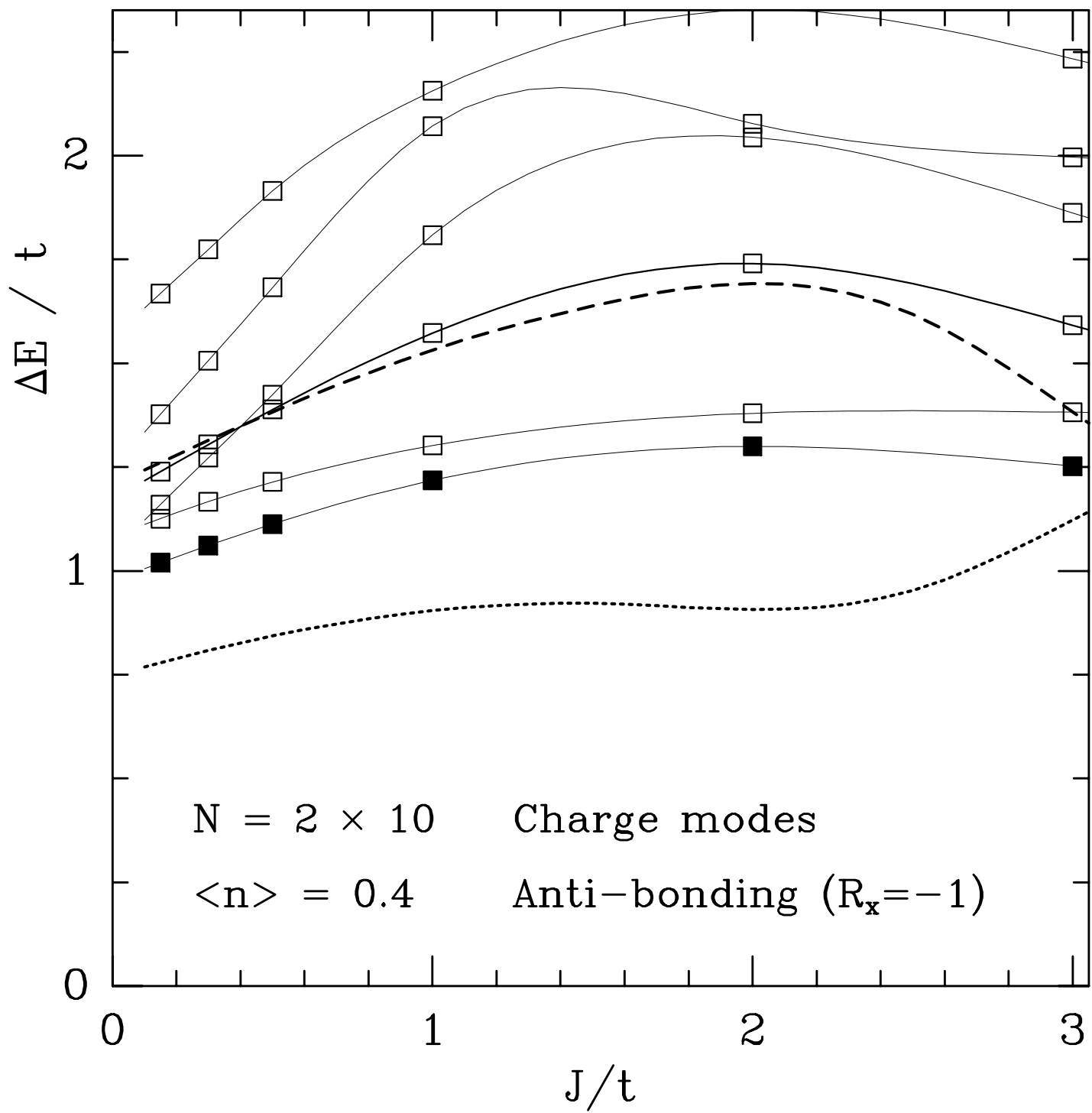

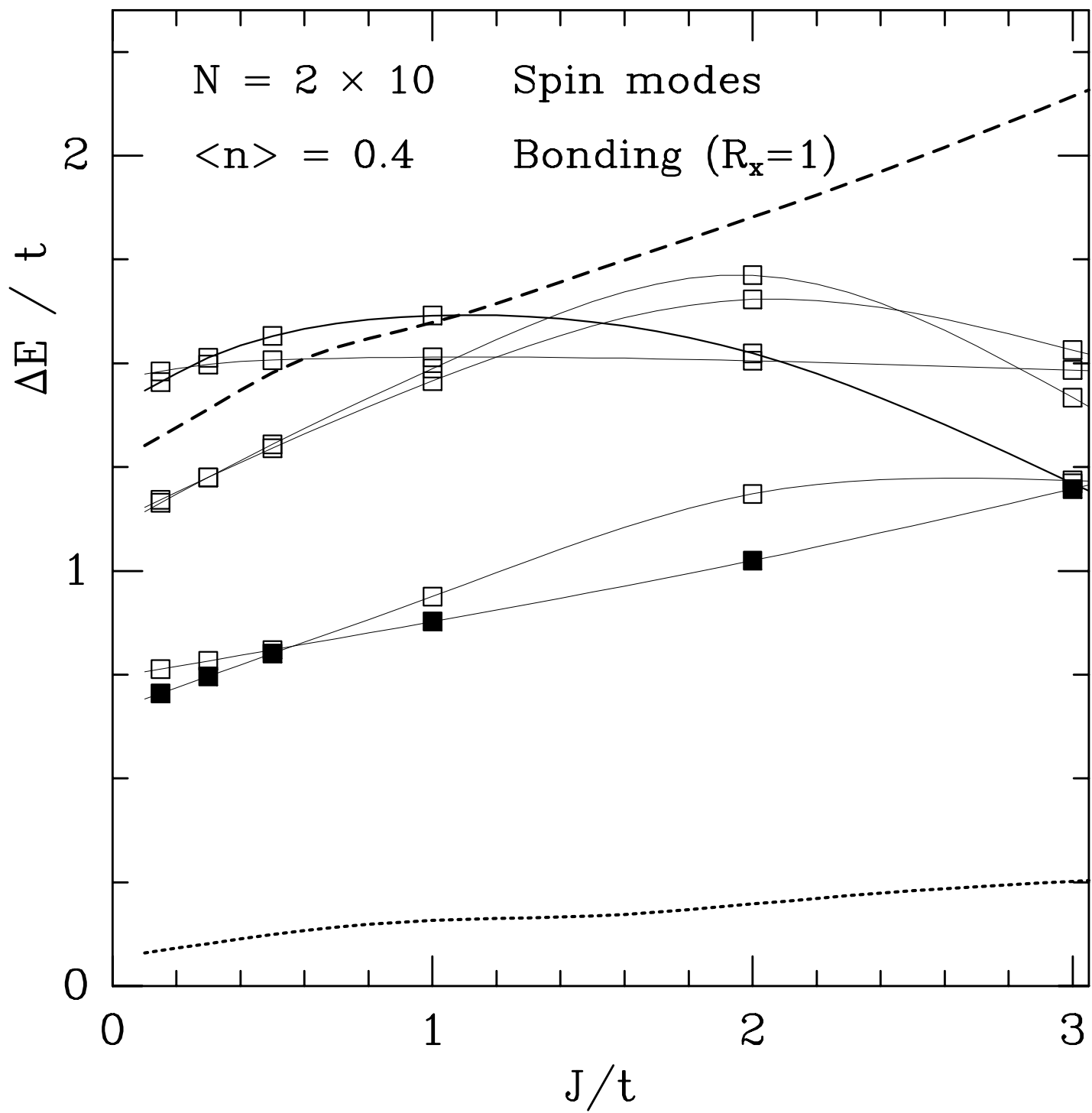

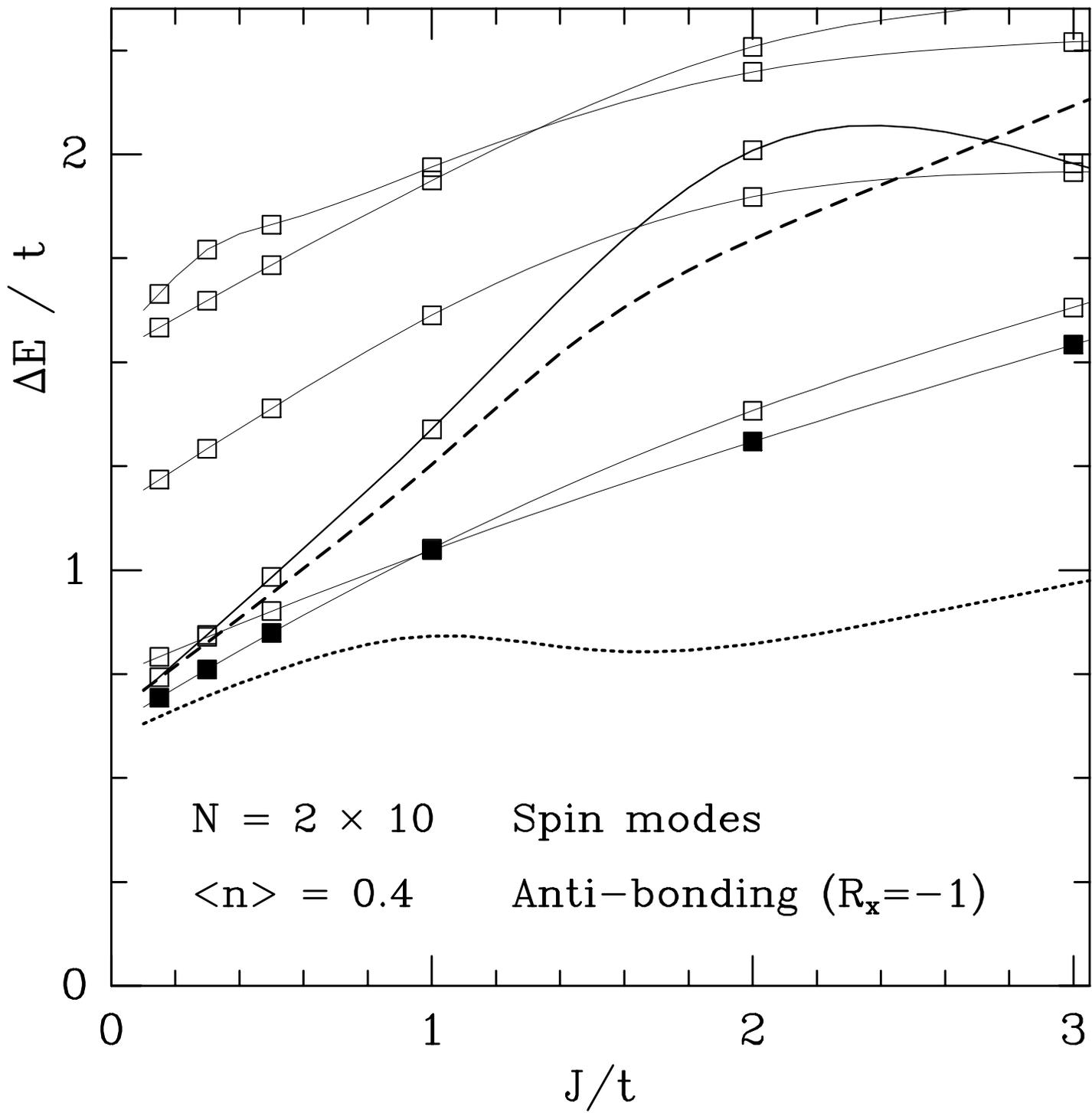

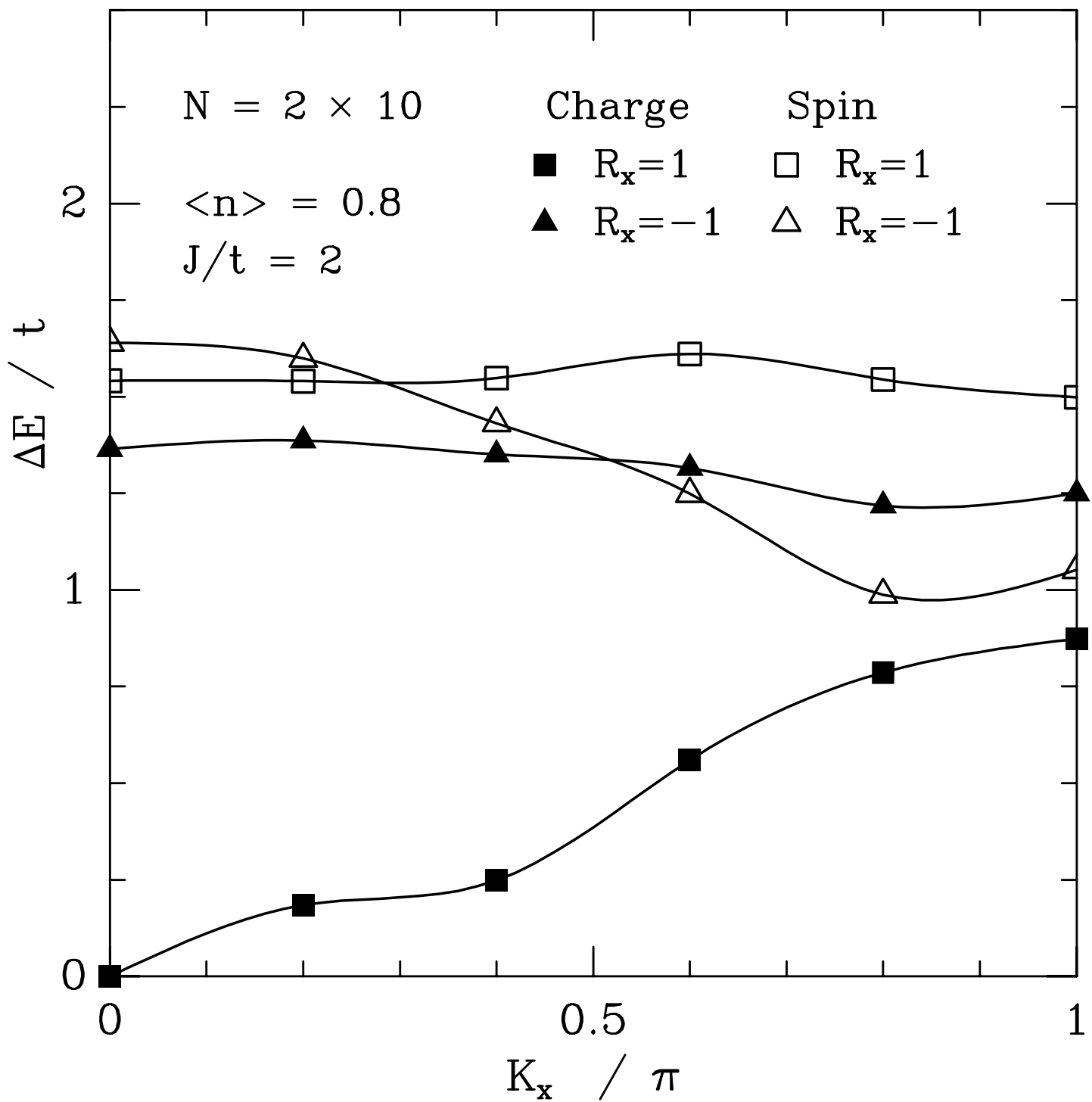

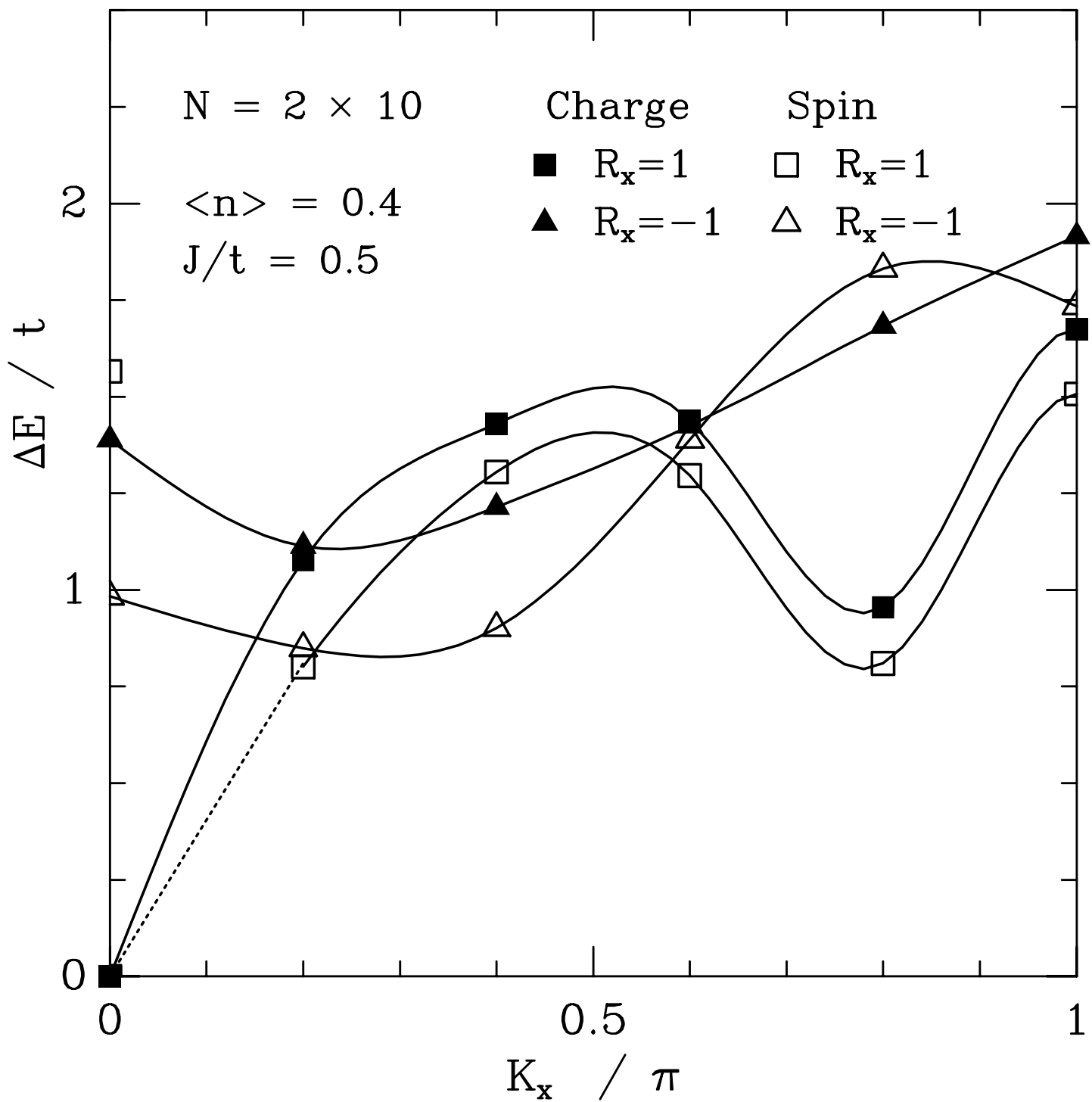

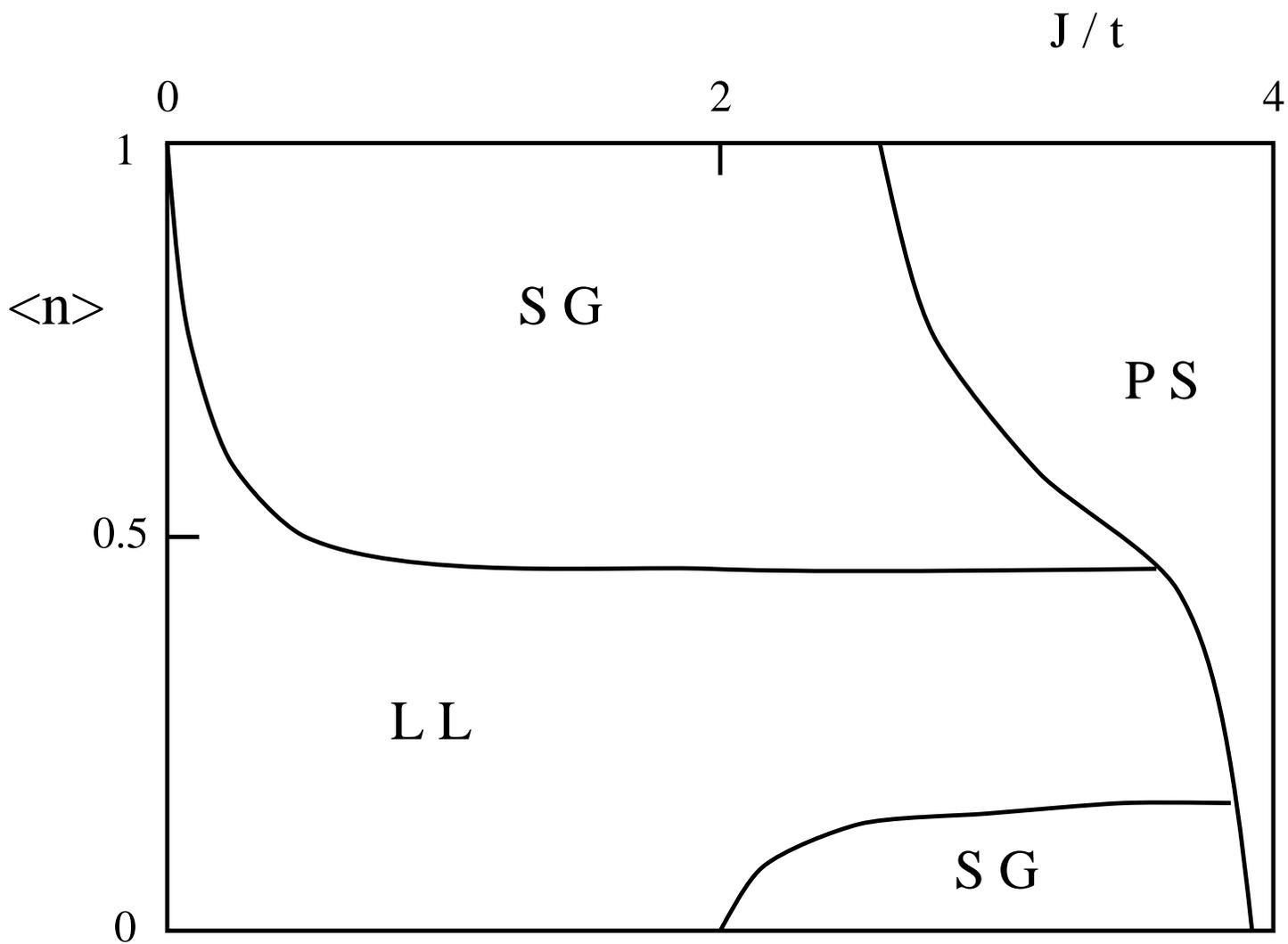

# Spin and charge modes of the $t$–$J$ ladder

D. Poilblanc[1,2], D.J. Scalapino[2] and W. Hanke[3,2]

[1] *Lab. de Physique Quantique, Université Paul Sabatier, 31062 Toulouse, France*
[2] *Department of Physics, University of California at Santa Barbara, Santa Barbara, CA 93106*
[3] *Institute for Physics, University of Würzburg, D-8700 Würzburg, Germany*

(Received)

The spin and charge excitations of the t–J ladder are studied by exact diagonalization techniques for several electron densities. The various modes are classified according to their spin (singlet or triplet excitations) and their parity under a reflection with respect to the symmetry axis along the chains and a finite size scaling of the related gaps is performed. At low doping formation of hole pairs leads to a spin gap for all $J/t$ ratios. This phase is characterized by (at least) one vanishing energy mode *only* in the charge bonding channel when $q_x \to 0$ consistent with dominant superconducting correlations. At larger doping the spin gap disappears. Although the anti-bonding modes remain gapped, low energy $q_x \sim 0$ and $q_x \sim 2k_F$ spin and charge bonding modes are found consistent with a single band Luttinger scenario. At sufficient low electron density and above a critical value of J/t we also expect another phase of electron pairs with gapped spin excitations.

The physical properties of t–J ladders are of interest both because of their possible realization in various materials such as $(VO)_2P_2O_7$[1] and $SrCu_2O_3$[2] and because they offer new insights into the nature of magnetic, charge density, and pairing correlations in strongly coupled electron systems. Here we report results for the charge and spin excitation spectrum of a doped isotropic t–J ladder obtained from exact diagonalization calculations on $2\times 10$ and $2\times 5$ clusters. Based on these results, we discuss a possible J/t-$\langle n \rangle$ ground state phase diagram.

The $t$-$J$ Hamiltonian on the $2 \times L$ ladder is defined as,

$$\begin{aligned}
\mathcal{H} = & J' \sum_j (\mathbf{S}_{j,1} \cdot \mathbf{S}_{j,2} - \tfrac{1}{4} n_{j,1} n_{j,2}) \\
& + J \sum_{\beta,j} (\mathbf{S}_{j,\beta} \cdot \mathbf{S}_{j+1,\beta} - \tfrac{1}{4} n_{j,\beta} n_{j+1,\beta}) \\
& - t \sum_{j,\beta,s} P_G (c^\dagger_{j,\beta;s} c_{j+1,\beta;s} + \mathrm{H.c.}) P_G \\
& - t' \sum_{j,s} P_G (c^\dagger_{j,1;s} c_{j,2;s} + \mathrm{H.c.}) P_G ,
\end{aligned} \quad (1)$$

where most notations are standard. $\beta$ $(=1,2)$ labels the two legs of the ladder (oriented along the $x$-axis) while $j$ is a rung index $(j=1,...,L)$. In the following we shall consider the isotropic case where the intra-ladder (along $x$) couplings J and $t$ are equal to the the inter-ladder (along $y$) couplings $J'$ and $t'$ respectively. However, to clarify the discussion, we have explicitly separated the intra- and inter-ladder parts in the Hamiltonian.

The half-filled case (Heisenberg ladder) is well understood.[3–6] When $J=0$, $\mathcal{H}$ reduces to isolated rungs and the spin excitations can be constructed by exciting singlet rungs to triplets each costing an excitation energy, $J'$. Once the intra-ladder coupling $J$ on the rungs is turned on, the triplets can propagate, form a coherent band (with a minimum energy at momentum $K_x=\pi$)[5], and the spin gap is reduced.

Upon lightly doping the t–J ladder the formation of hole pairs with persistence of the spin gap has been established by various numerical[7,8] or analytical techniques[9]. The nature of pairing is particularly evident when $J'$ is the largest energy scale since, in this case, it is energetically favorable for each hole pair to lie on a single rung to minimize the energy cost due to the destruction of singlet bonds. Interestingly enough pairing continues to exist in the isotropic case $J = J'$, even at small ratio J/t or in the regime of intermediate couplings described by the Hubbard model[10]. These bound pairs were also found to be quite robust under increasing the number of coupled chains.[11] However, the evolution of the spin gap under further doping is still unclear. At low *electron* density a phase of *electron* pairs exists above a critical ratio J/t where the two-electron bound state becomes stable. Therefore, under hole doping, a transition from a hole pair phase to an electron pair phase must occur. Since at small electron or hole concentrations the spin gaps of the paired phases are continuously suppressed by electron or hole doping respectively there is a possibility that an intermediate phase with gapless spin excitations may exist.

The goal of this paper is then to identify the various possible phases by determining the exact nature of their low energy excitations. A semi-quantitative derivation of the phase diagram can then be attempted. The low energy modes are first characterized by their spin. Singlet and triplet excitations correspond to charge and spin modes respectively. It is also very usefull to consider the parity of the excited states under a reflection with respect to the symmetry axis of the ladder along the direction of the chains. Even ($R_x = 1$) or odd ($R_x = -1$) excitations correspond to bonding or anti-bonding modes respectively.[12] Lastly, the total momentum $K_x$ along the ladder is considered in order to determine the form of the dispersion relation of each mode.

Before going further it is intructive to consider the excitation spectra for free fermions on a ladder.[13] At low enough doping the fermi level intersects the anti-bonding and bonding branches (split by an energy $2t' = 2t$) at $\pm k_{F,1}$ and $\pm k_{F,2}$ respectively. Excitations of bonding (anti-bonding) nature correspond to intra-band (inter-band) transitions. The bonding excitation spectrum consists then of four gapless modes (both in the spin and charge sectors) at momenta $q_x \sim 0$ (2 modes), $2k_{F,1}$ and $2k_{F,2}$. In the anti-bonding sector vanishing energy modes appear only at finite momenta $\pm(k_{F,1} \pm k_{F,2})$. It is already clear that this simple picture breaks down in the t-J ladder in the region where the spin excitations are gapped. When the hole pair concentration is very small one can clearly identify two different kinds of spin excitations;[8] (i) a triplet excitation of a singlet rung away from the hole pairs (typical of the half-filled case) with $R_x = -1$ and (ii) an excitation of a hole pair into 2 propagating quasi-particles with $R_x = 1$. Crudely, these two processes involve a *finite* excitation energy proportional to $J'$. Note that, with increasing doping, these bonding and anti-bonding spin modes loose their simple physical interpretation. At intermediate electron densities when the Fermi level only crosses the anti-bonding branch (for $\langle n \rangle < 0.5$) we shall see that the weakly interacting picture acquires nevertheless some relevance. In that case the anti-bonding (interband) modes are all gapped and, as far as the low energy modes are concerned, one only deals with a single band.

As seen above, numerical techniques have already proven to be very successful in studying coupled chains. Exact diagonalizations (ED) methods are particularly well adapted to the investigation of the low energy modes since implementation of quantum numbers and symmetries is rather straightforward. The following study is based on exact diagonalisations of $2 \times 5$ and $2 \times 10$ double rings at intermediate electron densities $\langle n \rangle = 0.4$ and $\langle n \rangle = 0.8$ corresponding to the less understood region of the phase diagram. Note that the electron number is always $N_e = 4p$ to guaranty that the antiferromagnetic correlations are not frustrated when one goes around each chain. Special care is also needed when choosing the boundary conditions (BC). Finite size corrections are reduced by using BC such that a non-interacting Fermi sea (obtained by turning off the interaction) would form a closed shell. This is achieved with anti-periodic (periodic) BC for $\langle n \rangle = 0.4$ ($\langle n \rangle = 0.8$). The various excitation modes are then straightforwardly obtained by calculating the GS energy in each symmetry sector with the Lanczos algorithm.

Charge and spin modes are shown in Figs. 1(a-d) for $\langle n \rangle = 0.8$ as a function of J/t. Data for all possible momenta are displayed but the precise behavior of the excitation energies vs $K_x$ will be discussed later on. Let us rather concentrate first on the energy gaps defined as the smallest excitation energies for each set of quantum numbers (ie the 4 possible bonding and anti-bonding charge and spin excitations) irrespective of momentum. In Figs. 1 the lowest excitation energy (gap) of the $2 \times 10$ cluster at each J/t value is indicated by a full square. The long-dashed line passes through the equivalent gap for the $2 \times 5$ cluster at the same electron density. Under doping the system becomes metallic and general hydrodynamic arguments predict long wavelength zero energy charge fluctuations. We then expect the finite size gap associated with the bonding charge mode to scale like $1/L$ (critical behavior)[14]. The short-dashed line in Fig. 1 corresponds to a simple $a + b/L$ extrapolation of the $2 \times 5$ and $2 \times 10$ data (where it is not shown the extrapolation gave a small negative number). As seen in Fig. 1(a) this procedure was found to give a very small (or even negative) gap in agreement with the existence of a zero energy mode in the $R_x = 1$ (bonding) charge channel. When we examine the $K_x$ momentum dependence we will see that this corresponds to the long wavelength $K_x \to 0$ mode. We expect that this type of extrapolation will in general give a *lower bound* for the gap[14] since, in the true asymptotic regime (ie when the system size exceeds the correlation length), the finite size corrections change to an exponential behavior with $1/L$. Figs. 1(b) and (d) strongly suggests the existence of a gap in the anti-bonding spin and charge modes. Fig. 1(c) shows that a gap also exists above a small critical value (our data give $J/t|_c \sim 0.4$ but $J/t|_c$ could be much smaller). We also find that all the gaps scale almost linearly with J/t. Above $J/t|_c$, the relevant physical picture is then consistent with a gas of hole pairs propagating along the chain direction. Below $J/t|_c$ it is likely that the hole pairs dissociate into two quasi-particles (QP) of opposite spins. In this phase a zero energy bonding spin excitation can be simply obtained by flipping the spin of one of the quasi-particles. However, an anti-bonding excitation requires the transfer of an anti-bonding QP to a bonding QP or the excitation of a singlet rung into a triplet. Both of these processes require a finite energy cost in agreement with Figs. 1(b) and (d).

Hole pairs are also expected to become unstable with increasing doping. Similar data as Fig. 1 for the charge and the spin excitations are shown in Figs. 2(a-d) for a smaller electron density $\langle n \rangle = 0.4$. Again Fig. 2(a) is consistent with the existence of zero energy (long wavelength) charge fluctuations characteristic of a metallic behavior. Figs. 2(b) and (d) show the existence of a sizable gap in the spectra of the anti-bonding modes. This is physically clear in the picture of the non- (or weakly) interacting fermi gas where the Fermi level only intersects the lower anti-bonding band (for $\langle n \rangle < 0.5$). Electron-hole transitions from the lower band to the upper band are then gapped and we expect the gaps to depend weakly on J in agreement with Figs. 2(b) and (d). The interpretation of Fig. 2(c) is more subjective. The extrapolated gap remains small and, contrary to Fig. 1(c), is weakly J dependent. This suggests that the spin gap vanishes in this region of the phase diagram.

More insights on the nature of the two spin gap and gapless phases found previously can be obtained from the dispersions of the four modes (ie the excitation energies vs $K_x$). The latter are shown in Fig. 3(a) and (b) for $\langle n \rangle = 0.8$, $J/t = 2$ and $\langle n \rangle = 0.4$, $J/t = 0.5$ respectively. In Fig. 3(a) the low frequency charge fluctuations are clearly separated from the three other modes at higher energy. Note also that there is no indication for any *finite momentum* zero mode contrary to the non-interacting case. Nevertheless, an investigation of the dynamical charge correlation function would clearly be useful to study the *number* of $R_x = 1$, $K_x \to 0$ charge modes (other zero bonding modes with larger charge velocities could also exists). Fig. 3(b) is more intricate. The bonding charge and spin excitation branches are very similar and seem to exhibit $q_x \sim 0$ and $q_x \sim 2k_F$ low energy modes while both anti-bonding modes are gapped. This is reminiscent of a single band Luttinger Liquid[15].

We conclude this paper with a brief discussion of a possible phase diagram which is sketched in Fig. 4. At large J/t the system separates into a hole-free and a hole-rich phase. The phase separation line at large J/t has been computed by Tsunetsugu et al.[8] from the compressibility. At smaller J/t ratios, when the electron density is reduced our data suggest a transition from a spin gap phase to a Luttinger-like phase[15] with gapless spin and charge excitations. At small electron density another paired phase must also appear due to the existence of a two-particle boundstate above $J/t|_c = 2$. Other $2p$-particle ($p > 1$) boundstates could also become stable at larger J/t in this region.

In the spin gap phase, we find evidence that there is only one gapless charge density mode which occurs for $K_x \to 0$ in the $R_x = 1$ (bonding) channel. This is consitent with the power law behavior of the singlet rung-rung pairing correlations found in recent numerical renormalization group calculations (NRG).[10] Furthermore, the gap in the charge excitation spectrum at $2k_F$ is also consistent with the NRG results which fail to find the conjugate, Luther-Emery,[16] power law decay of the $2k_F$ charge density correlations. It would appear that these conjugate correlations must developp in another channel besides the particle-hole charge density excitations studied here.

We thank M.P.A. Fisher for his insightful discussions regarding the relationship of gapless charge modes and pairing correlations. We gratefully acknowledge many useful discussions and comments from M. Imada, M. Luchini and F. Mila. *Laboratoire de Physique Quantique, Toulouse is Unité de Recherche Associé au CNRS No 505*. DP and DJS acknowledge support from the EEC Human Capital and Mobility program under Grant No. CHRX-CT93-0332 and the National Science Foundation under Grant No. DMR92-25027 respectively. We also thank IDRIS (Orsay) for allocation of CPU time on the C94 and C98 CRAY supercomputers.

[1] D.C. Johnston et al., *Phys. Rev. B* **35**, 219 (1987).
[2] Z. Hiroi et al., *Physica C* **185-189**, 523 (1991); M. Takano et al., *JJAP Series* **7**, 3 (1992).
[3] T.M. Rice, S. Gopalan, and M. Sigrist, *Europhys. Lett.* **23**, 445 (1993).
[4] T. Barnes, E. Dagotto, J. Riera, and E. Swanson, *Phys. Rev. B* **47**, 3196 (1993).
[5] S. Gopalan, T.M. Rice and M. Sigrist, *Phys. Rev. B* **49**, 8901 (1994).
[6] S.R. White, R.M. Noack, and D.J. Scalapino, SISSA preprint cond-mat/9403042.
[7] E. Dagotto, J. Riera, and D.J. Scalapino, *Phys. Rev. B* **45**, 5744 (1992).
[8] H. Tsunetsugu, M. Troyer, and T.M. Rice, *Phys. Rev. B*, in press (1994); see also H. Tsunetsugu, M. Troyer, and T.M. Rice, SISSA preprint cond-mat/9408024.
[9] M. Sigrist, T.M. Rice and F.C. Zhang, *Phys. Rev. B*, in press (1994).
[10] R.M. Noack, S.R. White and D.J. Scalapino, SISSA preprint cond-mat/9409065; the Renormalization Group applied to fermion lattice models is introduced in S.R. White, *Phys. Rev. Lett.* **69**, 2863 (1992).
[11] D. Poilblanc, H. Tsunetsugu and T.M. Rice, SISSA preprint cond-mat/9405045.
[12] One can alternatively view the bonding and anti-bonding states as carrying a momentum $K_y = 0$ or $\pi$ respectively in the transverse direction.
[13] Stricly speaking spin and charge modes are degenerate in this limit. A small Hubbard U introduced at the crude RPA level removes this degeneracy without changing the qualitative conclusions.
[14] F. Mila and X. Zotos, *Europhys. Lett.*, **24**, 133 (1993); see also G. Sponken, R. Jullien and M. Avignon, *Phys. Rev. B* **24**, 5356 (1981).
[15] J.M. Luttinger, *J. math. Phys.* **4**, 1154 (1963); for a review see J. Sólyom, *Advances in Physics* **28**, 201-303 (1979).
[16] A. Luther and V.J. Emery, *Phys. Rev. Lett.* **33**, 589 (1974).

FIG. 1. Charge ((a) and (b)) and spin ((c) and (d)) excitations of a $2 \times 10$ ladder at electron filling $\langle n \rangle = 0.8$ as a function of the ratio J/t. The bonding (anti-bonding) modes correspond to the figures (a) and (c) ((b) and (d)). The energy reference is given by the GS energy and periodic boundary conditions in the chain direction are used. Data for all possible momenta ($K_x = 2\pi n/L$, n integer) are shown (open and full squares) and the various data sets corresponding to the same momentum are connected by a continuous thin line. The lowest excitation energy (gap) is indicated by full squares for the $2 \times 10$ cluster and is also shown for a smaller $2 \times 5$ cluster (at the same electron density) by a long-dashed line. The short-dashed line corresponds to a tentative $1/L$-extrapolation of this finite size gap.

FIG. 2. Same as Fig. 1 for an electron density of $\langle n \rangle = 0.4$ and with anti-periodic boundary conditions in the direction of the chains.

FIG. 3. Spin and charge excitation modes of a $2 \times 10$ ladder versus momentum $q_x$ (in unit of $\pi$). The quantum numbers associated to the various symbols are shown on the plots. (a) and (b) correspond to $\langle n \rangle = 0.8$ and $J/t = 2$ and to $\langle n \rangle = 0.4$ and $J/t = 0.5$ respectively.

FIG. 4. Tentative schematic picture of the $J/t - \langle n \rangle$ phase diagram. The boundary with the phase separated region is taken from Ref. 8